# Concurrent probing of electron-lattice dephasing induced by photoexcitation in 1T-TaSeTe using ultrafast electron diffraction

Jun Li[1], Junjie Li[1], Kai Sun[2], Lijun Wu[1], Renkai Li[3], Jie Yang[3], Xiaozhe Shen[3], Xijie Wang[3], Huixia Luo[4], Robert J. Cava[4], Ian K. Robinson[1,5], Xilian Jin[1,6], Weiguo Yin[1], Yimei Zhu[1], Jing Tao[1, *]

[1]Condensed Matter Physics & Materials Science Department, Brookhaven National Laboratory, Upton, NY 11973, USA

[2]Department of Physics, University of Michigan, Ann Arbor, MI 48109, USA

[3]SLAC National Accelerator Laboratory, Menlo Park, CA, USA.

[4]Department of Chemistry, Princeton University, Princeton, NJ 08544, USA

[5]London Centre for Nanotechnology, University College, London WC1E 6BT, UK

[6]College of Physics, Jilin University, Changchun, Jilin 130012, P.R. China

It has been technically challenging to concurrently probe the electrons and the lattices in materials during non-equilibrium processes, allowing their correlations to be determined. Here, in a single set of ultrafast electron diffraction patterns taken on the charge-density-wave (CDW) material 1T-TaSeTe, we discover a temporal shift in the diffraction intensity measurements as a function of scattering angle. With the help of dynamic models and theoretical calculations, we show that the ultrafast electrons probe both the valence-electron and lattice dynamic processes, resulting in the temporal shift measurements. Our results demonstrate unambiguously that the CDW is not merely a result of the periodic lattice deformation ever-present in 1T-TaSeTe but has significant electronic origin. This method demonstrates a novel approach for studying many quantum effects that arise from electron-lattice dephasing in molecules and crystals for next-generation devices.

Observation of the incoherent movements of electrons and lattice, i.e., electron-lattice dephasing in excited states is of fundamental importance to understand charge-lattice interactions [1-5]. Recently, the rapid development of ultrafast methods offers the opportunity to trace dynamics in structures for characterizing the charge-lattice interactions induced by photoexcitations [1-3, 6-12]. Many ultrafast studies have focused on CDW materials [1, 7-12] that are of great interest due to their intimate relation to a variety of captivating electronic properties including metal-insulator transition and superconductivity [4, 5, 13-15]. The symmetry-broken states of CDW are depicted by real-space charge-density re-distributions, which often result in a periodic lattice distortion (PLD) at equilibrium [16], or vice versa. It is this "electron dichotomy" reflecting distinct electronic and lattice contributions to the CDW that gives rise to much of the ongoing debates about the nature and origin of CDWs in various systems. One of the most compelling challenges in studying CDW materials is to probe the dynamics of charge states and the lattice distortions concurrently because only the co-evolution can yield the correct understanding of the driving mechanisms [1, 6, 7]. Limited by technical difficulties, however, the electron dynamics and lattice evolution in a material during non-equilibrium processes are commonly investigated by discrete methods, for instance, angle resolved photon emission spectroscopy for electron dynamics [17, 18] and x-ray diffraction for lattice dynamics [19]. Correlating the observations of the different experimental methods requires careful synchronization, however, is often an insurmountable problem due to the distinct nature of the probes and the different experimental and material conditions employed.

Electron diffraction techniques have been developed for nearly a century, with intrinsic advantages compared to other scattering tools [20] due to the high electron scattering cross-sections. Unlike for X-rays, which interact with all the electrons surrounding an atom, and are therefore sensitive to atom positions, electrons interact with the electrostatic potential of an atom - its positively charged nucleus screened by its negatively charged electron cloud [21-23]. Thus, the scattering amplitude of an atom for incident electrons at small scattering angles is determined mainly by an atom's valence charge, rather than by the total density of electrons (see Fig.1(a)). In other words, the electron scattering atomic form factor at small scattering angles is strongly influenced by the valence electron distribution of atoms and charge transfer in a crystal. This is particularly advantageous for crystals with large unit cells that have reflections at small scattering angles [21]. In contrast, in a complementary fashion, electrons scattered to high scattering angles are extremely sensitive to subtle changes of atom positions or atomic motion.

Inspired by this feature of the scattering mechanism, we analyze the intensity of super-lattice reflections (SLRs) that correspond to the CDW superstructure in the single-crystalline layered transition metal dichalcogenide 1T-TaSeTe using MeV ultrafast electron diffraction (UED) [24, 25], taking advantage of its sensitivity to the valence charge and atomic displacements, which dominate at different scattering angles. The 1T-TaSeTe single crystals were grown by chemical vapor transport with iodine as a transport agent [26]. In fact, the PLD in our samples exists at all the measureable temperatures until they melt. This raises a serious question about whether the

CDW in this material is truly of some electronic origin or merely a result of the ever-present lattice deformation. Understanding this exeme case is instrumental to resolving the aforementioned ongoing debate about the nature and origin of CDWs in various systems. The UED experiments were performed on the 3.5-MeV-UED setup at the Stanford Linear Accelerator Laboratory excited by a laser pulse with a photon energy of 1.55 eV (center wavelength of 800 nm) and a fluence of 3.5 mJ/cm$^2$ [27]. This classic layered 1T dichalcogenide has a ramdom Te-Se distribution on the dichalcogenide lattice site [26]; the Te/Se mixture is required to stabilize the 1T structure. The measurement temperature is 26 K to minimize the thermal atomic motion and diffuse scattering background. A portion of the UED pattern (at t = -2 ps; before pumping) is shown in Fig. 1(b), with nine sets of the SLRs and their positions in reciprocal space as a function of scattering angles marked by the dashed line (see supplementary information for a full UED pattern in Fig. S1 and all the experimental data in Fig. S2). The subscripts 1, 2 and 3 correspond to indexes ($hkl$100), ($hkl$010) and ($hkl$0$\bar{1}$0), respectively, owing to the triple-q states in the CDW [26, 27]. Normalized intensities (normalized to the averaged intensity before t = 0) as a function of time delay from three selected SLRs are shown in Fig. 1(c). Through quantitative intensity analysis we determine the atomic positions during the lattice evolution upon photoexcitation [27]. The SLR intensities as a function of time are simulated and shown in supplementary Fig. S4. The simulation is consistent with the experiment, revealing that there is a rapid drop of the normalized intensities, which reach a minimum at a few hundred femtoseconds, then recover on a slower time scale. We use a two-exponential function to fit the SLR intensities in Fig. 1(c) which well describes the experimental observations.

Compared with the reconstructed lattice evolution using the same set of UED data [27], we notice that at the time when the SLR intensities reach the minimum (the cusp), both the Ta and Se/Te atoms depart furthest from their PLD state with distorted positions (i.e., before pumping). The temporal characteristics in the lattice evolution, particularly the time of the cusp $t_c$, should be equally reflected by all the SLRs suggested by the diffraction simulations of all the SLRs that $t_c$ is the same given the same set of the atomic displacements through the dynamics (see simulated intensity variations in Fig. S4 and Fig. S5 using both kinematic and dynamic electron scattering simulations). However, unexpectedly we observe an intriguing difference in values of $t_c$ measured for the SLRs at different scattering angles. To better visualize the shift of $t_c$, we renormalized the intensity plots (Fig.2(a)). The intensity renormalization affects the time constants in the two-exponential curve fitting but it does not affect the value of $t_c$. In Fig. 2(a), the curve fits of the intensity variations of all nine SLRs show a shift of $t_c$ toward higher values as the scattering vector length s increases (also see inset). The measured $t_c$ vs. s behavior is plotted in Fig. 2(b) (black dots), indicating that the value of $t_c$ changes from ~ 0.48 ps for the (100$_1$) reflection to ~ 0.88 ps for the (400$_1$) reflection.

Since the evolution of atomic displacements in the lattice do not induce the measured shift in $t_c$, we move further to examine the lattice vibration effect on the SLRs, which are sensitive to scattering angle, on the measured $t_c$ values. The normalized SLRs intensity at time $t'$ can be expressed by $\frac{I(t')}{I(t_0)} = \frac{I(u')}{I(u_0)} e^{-2B(t')s^2}$, where $t_0$ is time zero, $u'$ the lattice distortion at $t'$ and $u_0$ the lattice distortion at $t_0$, $B(t')$ is the effective Debye-Waller (D-W) factor at $t'$ assuming isotropic and identical for all the atoms, and $s = sin\theta/\lambda$ ($\theta$ being the half of scattering angle and $\lambda$ is the electron wavelength) as the scattering vector [20, 22]. It is widely accepted during the warm up $B(t')$ can be expressed by $B(t') = a(1 - e^{-t'/\tau_{DW}})$ assuming the lattice has negligible vibration at time zero, where $\tau_{DW}$ is the time constant for the change in the lattice vibration and $a$ describes the value of $B(t')$ where it reaches saturation [28-30]. By applying the D-W term to the intensities of the SLRs, we find that the $t_c$ in the intensity variation can be shifted to higher values, depending on the parameters $a$ and $\tau_{DW}$ in the $B(t')$ expression. According to the intensity expression, we measured $\frac{I(t')}{I(t_0)}$ and calculated the $\frac{I(u')}{I(u_0)}$ for all nine SLRs at $t' = 6$ ps when the system reaches to a quasi-stable condition. Then the parameter $a$ is determined to be ~ 0.134 (Å$^2$) by the slope of the plot of $\ln\left[\frac{I(t')}{I(t_0)}/\frac{I(u')}{I(u_0)}\right]$ vs. $s^2$ (see Fig. S6). While it is hard to accurately determine $\tau_{DW}$, we find that the D-W term has a maximum effect on the $t_c$ value (i.e., $t_c$ exhibits the largest shif) when $\tau_{DW}$ ~ 0.9 ps (see Fig. S6). Given all the above considerations in measurements, we remove the lattice vibration effect on the intensity variations measured from the nine SLRs by dividing the raw data by $e^{-2B(t')s^2}$, in which $B(t') = 0.134(1 - e^{-t'/0.9})$, and fit the processed data again. The corrected values of $t_c$ are plotted as the red dots in Fig. 2(b), with the correction as scattering-angle dependent. Note that the correction for the D-W term is the upper limit of the effect in this case for the reason stated above. Therefore, the lattice behavior, including the atomic displacements and the lattice vibrations, cannot be responsible for the $t_c$ shift that is unambiguously observed in the experiment.

To explore the origin of the $t_c$ shift, we employ theoretical models to explain the dynamic behavior of the system. The electrons are excited abruptly by the pumping photons, and the excitation and relaxation process takes place within a few femtoseconds [6, 31, 32]. Assuming that an electronic order parameter $\eta$ in the material relaxes in an exponential decay from the excited states, then

$$\eta(t) = \eta_f + (\eta_i - \eta_f)e^{-t/\tau_\eta}, \qquad 1)$$

where $\tau_\eta$ is the time constant for the electron relaxation, and $\eta_i$ and $\eta_f$ are the initial and a semi-final stage (when the measurements reach meta-stable values after $t$ ~ 3 ps [27]) of the electronic order, respectively. On the other hand, the lattice order parameter $Q$ can be written in a dynamic equation

$$\frac{dQ}{dt} = \frac{Q - Q_f(\eta)}{\tau_Q}, \qquad 2)$$

where lattice order $Q_f$ is the final states (after 3 ps) is a function of $\eta$ instead of a simple time-independent constant and $\tau_Q$ describes how fast the lattice follows the change in the electronic order. Taking linear coupling between the lattice and electrons for simplicity, i.e., $Q_f(\eta) = \eta$, Eqn. 2) is $\frac{dQ}{dt} = \frac{Q - \left[\eta_f + (\eta_i - \eta_f)e^{-t/\tau_\eta}\right]}{\tau_Q}$, which has an analytical solution as follows.

$$Q(t) = e^{-t/\tau_Q} + \eta_f\left(1 - e^{-t/\tau_Q}\right) + \frac{(\eta_i - \eta_f)\tau_\eta}{\tau_Q - \tau_\eta}(e^{-t/\tau_Q} - e^{-t/\tau_\eta}) \qquad 3)$$

Eqn. 1) and 3) are plotted as the black and red (both bold) curves in Fig. 3 to represent the dynamics of electron and lattice, respectively, by setting $\eta_i = 0.6$, $\eta_f = 0.9$, $\tau_\eta = 0.4$ and $\tau_Q = 0.3$. Eqn. 3) provides a phenomenological interepration of the fitting functions that the two time-constants ($\tau_\eta$ and $\tau_Q$) are associated with the electron relaxation speed (even we measure the response of the lattice) and the speed that the lattice follows the change in electrons. The amplitudes of the two-exponential fittings, which have been often employed in UED data analysis [8, 12, 27, 33], can now be explicitly expressed by Eqn. 3). Indeed, to be comparable to the experimental results, $\tau_\eta$ and $\tau_Q$ need to have similar values (~ 0.3 ps in this case). This indicates that the relaxation time of the electronic order is not independent of its environment, but strongly coupled to the lattice dynamics. Such an implication is consistent with polaron-type behavior (i.e., the electron and lattice dynamics are intertwined) suggested by previous ultrafast-observations in a doped manganite [33]. Most interestingly, the theoretical plots clearly show that a mixture of the lattice order with the electronic order (see a linear combination of the black and red bold curves in Fig. 3) can explain the shift of the cusp of the curve in ultrafast-time regime. The more weight of electronic order in the mixed intensity, the faster the curve reaches its cusp. Imagining an electron beam that probes mainly the lattice dynamics with a certain portion of its diffraction intensity as being due to the electron dynamics, the intensity variation would be identical to the curves depicted in Fig. 3. In comparison to the experimental findings in Fig. 2(b) and based on the scattering principles illustrated in Fig. 1(a), we interpret the shift of measured $t_c$ in Fig. 2(b) as arising from the co-evolution of the lattice and the electron dynamics, which are both reflected in the diffraction intensity variations. Even a few percent of electron contribution to the total intensity yields a shift of $t_c$ for the data at small scattering vector length $s$, while at higher scattering angles (larger $s$), the value of $t_c$ is predominantly dictated by the lattice dynamics because the electron contribution is nearly zero.

The reflection of the electron dynamics in the UED measurements as a function of scattering vector are substantiated by density-functional-theory (DFT) calculations TaSe$_2$ in the 1T structure. The normal state, with a high-symmetry electron/lattice arrangement (no CDW), and the CDW state, with symmetry breaking, were both calculated; their charge density distributions in real-space are illustrated in Fig. 4(a). Structure factors (and intensities $I_{total}$) using total charge for the nine SLRs were further derived from the calculated structures and charge density mapping for both x-ray and electron diffraction. In addition, structure factors (and intensities $I_{valence}$) using valence electrons only, that are identified to be 5$d$ and 6$s$ electrons for Ta atoms and 4$p$ electrons for Se atoms (note that other orbital electrons can also be involved in the photoexcitation), were calculated as well using the DFT results. The ratio of $I_{valence}$ to $I_{total}$ are plotted in Fig. 4(b) as a function of $s$ for both x-rays and electrons. It clearly shows that both techniques manifest a scattering-angle-dependent intensity variation and the weight of valence electrons in the total intensity is much higher in electron diffraction than that in x-ray, particularly at small scattering angles. Thus the (valence) electron dynamics can well be reflected in the temporal characteristics measured from UED. Note that the DFT calculations in Fig. 4 are from 3×3-type CDW structures based on the experimental observations. Moreover, similar DFT calculations and the derived intensities for electron and x-ray diffractions from the Star-of-David-type (with a $\sqrt{13} \times \sqrt{13}$ unit cell) CDW, which is the low-temperature state of 1T-TaSe$_2$, can be found in supplementary Fig. S7 with discussions. Both CDW patterns have in common that in each cluster the six nearest-neighbor Ta atoms of the central Ta atom moving towards the center. With the help of theoretical modeling and calculations, the results show that the lattice dynamics are driven by the change in the electronic structure in this material, which addresses the question of whether the origin of the superstructure originates in chemical order or electronic instability (i.e., a CDW) [4, 5, 26, 34].

Separating scattering contributions of valence electrons from a lattice of atomic nuclei and inner-shell electrons is of great importance but not a trivial task with diffraction. Because incident electrons interact with electrostatic potentials of the sample, i.e., atomic nuclei screened by the electron clouds, electron diffraction has a better capability, compared to x-ray techniques, to distinguish the contribution of valence electrons from the total scattering intensity [ref. 21, and quantitatively manifested in this work]. Our findings demonstrate a novel experimental approach to concurrently probing both lattice and electron dynamics at an ultrafast time scale because the lattice and electrons move incoherently with distinct dynamics, manifesting that 1T-TaSeTe is a bona fide CDW material in which the CDW is not merely a result of the ever-present PLD and suggesting that CDW in systems that otherwise exhibit a PLD phase transition be even more likely to have significant electronic origin.

In a broader scope, in many correlated electron systems the low-energy electrons near the Fermi level tend to self-partition into fast (itinerant) and slow (more localized) ones via various mechanisms. Well-known examples include the cuprates [37] and the iridates [38] in which the partition occurs in the momentum space as node and antinode regions, the iron-based superconductors in which the partition takes place via the orbital-selective Mott transition [39,40], and the orbital-selective Peierls transitions in CuIr$_2$S$_4$ spinel [41] and NaTiSi$_2$O$_6$ pyroxene [42]. Again, it is this "electron dichotomy" that gives rise to much of the ongoing debates about the nature and origin of the various unusual phenomena in those systems. As the electrons move fast or slow through the lattice, their couplings to the lattice are substantially different. Hence, the present technique is anticipated to provide novel insights into many quantum materials by temporally separating and ultimately quantifying electron-lattice coupling on its fundamental time scales.

**Acknowledgements**

Research was sponsored by DOE-BES Early Career Award Program and by DOE-BES under Contract DE-SC0012704. The work at University of Michigan was supported by NSF-EFMA-1741618. The UED experiment was performed at SLAC MeV-UED, which is supported in part by the DOE BES SUF Division Accelerator & Detector R&D program, the LCLS Facility, and SLAC under contract Nos. DE-AC02-05-CH11231 and DE-AC02-76SF00515. The work at Princeton was supported by Department of Energy, Division of Basic Energy Sciences, Grant DE-FG02-98ER45706. X.J. acknowledges the visiting scholarship of Brookhaven National Laboratory and the financial support of China Scholarship Council.

Electronic address: jtao@bnl.gov

1. Tao, Z. *et al*. Decoupling of structure and electronic phase transitions in $VO_2$. *Phys. Rev. Lett.* **109,** 166406 (2012).
2. Ishioka, K., Hase, M. & Kitajima, M. Ultrafast electron-phonon decoupling in graphite. *Phys. Rev. B* **77,** 121402 (2008).
3. Stern, M. J. *et al*. Mapping momentum-dependent electron-phonon coupling and nonequilibrium phonon dynamics with ultrafast electron diffuse scattering. *Phys. Rev. B* **97,** 165416 (2018).
4. Rossnagel, K., On the origin of charge-density waves in select layered transition-metal dichalcogenides. *J. Phys.: Condens. Matter* **23,** 213001 (2011).
5. Zhu, X., Cao, Y., Zhang, J., Plummer, E. W. & Guo, J. Classification of charge density waves based on their nature. *Proc. Natl. Acad. Sci. USA* **112,** 2367-2371 (2015).
6. Konstantinova, T. *et al*. Nonequilibrium electron and lattice dynamics of strongly correlated $Bi_2Sr_2CaCu_2O_{8+\delta}$ single crystals. *Sci. Adv.* **4,** eaap7427 (2018).
7. Otto, M. R. *et al.*, How optical excitation controls the structure and properties of vanadium dioxide. *PNAS* **116**, 450 (2019)
8. Eichberger, M. *et al*. Snapshots of cooperative atomic motions in the optical suppression of charge density waves. *Nature* **468,** 799-802 (2010).
9. Han, T.-R. T. *et al*. Exploration of metastability and hidden phases in correlated electron crystals visualized by femtosecond optical doping and electron crystallography. *Sci. Adv.* **1,** e1400173 (2015).
10. Zong A. *et al*. Ultrafast manipulation of mirror domain walls in a charge density wave. *Sci. Adv.* **4**, eaau5501 (2018).
11. Wall, S. *et al*. Ultrafast disordering of vanadium dimers in photoexcited $VO_2$. *Science* **362,** 525-526 (2018).
12. Zong, A. *et al.* Evidence for topological defects in a photoinduced phase transition. *Nat. Phys.* **15**, 27-31 (2019).
13. Wilson, J. A. & Yoffe, A. D. The transition metal dichalcogenides discussion and interpretation of the observed optical, electrical and structural properties. *Adv. Phys.* **18,** 193-335 (1969).

14. Pollak, R. A., Eastman, D. E., Himpsel, F. J., Heimann, P. & Reihl, B. 1T-TaS$_2$ charge-density-wave metal-insulator transition and Fermi-surface modification observed by photoemission. *Phys. Rev. B* **24,** 7435-7438 (1981).
15. Chang, J. *et al*. Direct observation of competition between superconductivity and charge density wave order in YBa$_2$Cu$_3$O$_{6.67}$. *Nature Phys.* **8,** 871-876 (2012).
16. Wilson, J. A., Di Salvo, F. J. & Mahajan, S. Charge-density waves and superlattices in the metallic layered transition metal dichalcogenides. *Adv. Phys.* **24,** 117-201 (1975).
17. Smallwood, C. L., Kaindl. R. A. & Lanzara, A. Ultrafast angle-resolved photoemission spectroscopy of quantum materials. *Europhys. Lett.* **115,** 27001 (2016).
18. Gerber, S. *et al.* Femtosecond electron-phonon lock-in by photoemission and x-ray free-electron laser. *Science* 357, 71 (2017).
19. Gaffney, K. J. & Chapman, H. N. Imaging atomic structure and dynamics with ultrafast X-ray scattering. *Science* **316,** 1444-1448 (2007).
20. Zuo, J. M. & Spence, J. C. H. *Advanced transmission electron microscopy* (Springer, New York, 2017).
21. Zheng, J.-C., Zhu, Y., Wu, L. & Davenport, J. W. On the sensitivity of electron and X-ray scattering factors to valance charge distributions. *J. Appl. Cryst.* **38,** 648-656 (2005).
22. Peng, L.-M. Electron atomic scattering factors and scattering potentials of crystal. *Micron* **30,** 625-648 (1999).
23. Stefanou, M., Saita, K., Shalashilin, D. V. & Kirrander, A. Comparison of ultrafast electron and X-ray diffraction – A computational study. *Chem. Phys. Lett.* **683,** 300-305 (2017).
24. Wang, X. J., Wu, Z., & Ihee, H. Femoto-seconds electron beam diffraction using photocathode RF gun. *Proc. 2003 Part Accel. Conf.* **1,** 420-422 (2003).
25. Weathersby, S. P. *et al*. Mega-electron-volt ultrafast electron diffraction at SLAC National Accelerator Laboratory. *Rev. Sci. Instrum.* **86,** 73702 (2015).
26. Luo, H. *et al*. Ploytypism, polymorphism, and superconductivity in TaSe$_{2-x}$Te$_x$. *Proc. Natl. Acad. Sci. USA* **112,** E1174-E1180 (2015).
27. http://arxiv.org/abs/1903.09911
28. Wei, L. *et al.* Dynamic diffraction effects and coherent breathing oscillations in ultrafast electron diffraction in layered 1T-TaSeTe. *Struct. Dyn.* **4,** 044012 (2017).
29. Carbone, F., Yang, D. S., Giannini, E. & Zewail, A. H. Direct role of structural dynamics in electron-lattice coupling of superconducting cuprates. *Proc. Natl. Acad. Sci. USA* **105,** 20161 (2008).
30. Harb, M. *et al.* Phonon-phonon interactions in photoexcited graphite studies by ultrafast electron diffraction. *Phys. Rev. B* **93,** 104104 (2016).
31. Tao, Z. *et al*. Direct time-domain observation of attosecond final-state lifetimes in photoemission from solids. *Science* **353,** 62-67 (2016).
32. Chen, J. K., Tzou, D. Y. & Beraun, J. E. A semiclassical two-temperature model for ultrafast laser heating. *Int. J. Heat Mass Tran.* **49,** 307-316 (2006).

33. Li, J. *et al.* Dichotomy in ultrafast atomic dynamics as direct evidence of polaron formation in manganites. *NPJ Quant. Mater.* **1,** 16026 (2016).
34. Liu, Y. *et al*. Nature of charge density waves and superconductivity in 1T-TaSe$_{2-x}$Te$_x$. *Phys. Rev. B* **94,** 045131 (2016).
35. Zhu, P. *et al*. Femtosecond time-resolved MeV electron diffraction. *New J. Phys.* **17,** 063004 (2015).
36. Midgley, P. A. & Eggeman, A. S. Precession electron diffraction – a topical review. *IUCrJ* **2,** 126-136 (2015).
37. M. R. Norman, H. Ding, M. Randeria, J. C. Campuzano, T. Yokoya, T. Takeuchi, T. Takahashi, T. Mochiku, K. Kadowaki, P. Guptasarma, D. G. Hinks. Destruction of the Fermi Surface in Underdoped High Tc Superconductors. *Nature* **392**, 157 (1998).
38. Y. K. Kim, O. Krupin, J. D. Denlinger, A. Bostwick, E. Rotenberg, Q. Zhao, J. F. Mitchell, J. W. Allen, B. J. Kim. Fermi arcs in a doped pseudospin-1/2 Heisenberg antiferromagnet. *Science* **345**, 187 (2014).
39. Luca de' Medici, S. R. Hassan, Massimo Capone, and Xi Dai. Orbital-Selective Mott Transition out of Band Degeneracy Lifting. *Phys. Rev. Lett.* **102**, 126401 (2009).
40. W.-G. Yin, C.-C. Lee, and W. Ku. Unified Picture for Magnetic Correlations in Iron-Based Superconductors. *Phys. Rev. Lett.* **105**, 107004 (2010).
41. E.S. Bozin, W.G. Yin, R.J. Koch, M. Abeykoon, Y.S. Hor, H. Zheng, H.C. Lei1, C. Petrovic, J.F. Mitchell & S.J.L. Billinge. Local orbital degeneracy lifting as a precursor to an orbital-selective Peierls transition. *Nat. Commun.* **10**:3638 (2019).
42. A. Feiguin, A. M. Tsvelik, Weiguo Yin, E. S. Bozin. Quantum liquid with strong orbital fluctuations: the case of a pyroxene family. *Phys. Rev. Lett.* **123**, 237204 (2019).

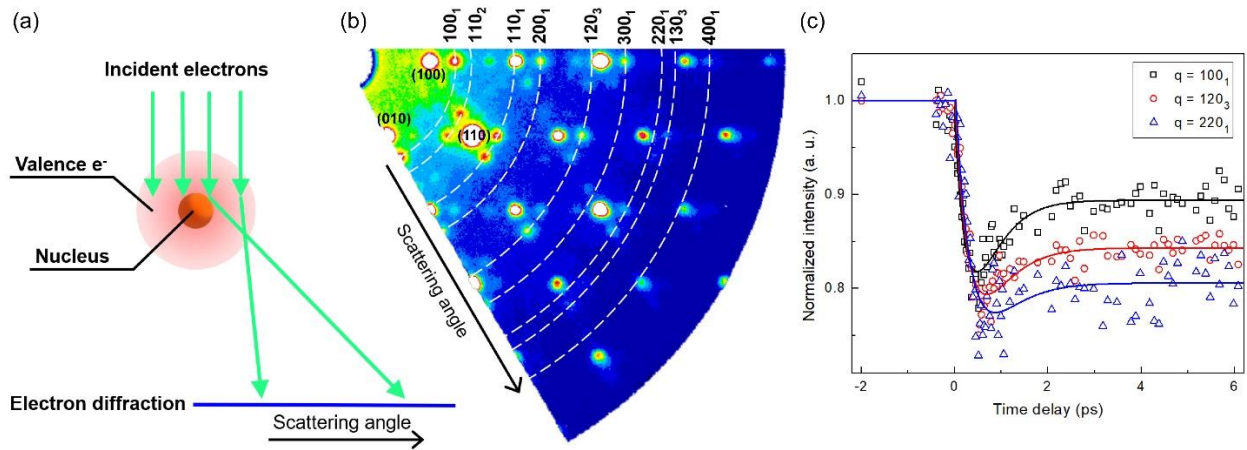

Figure 1. Scattering-angle-dependent dynamics measured via ultrafast electron diffraction (UED). (a) A schematic representation of electron scattering, illustrating that scattering from valence electrons is predominant at small scattering angles, while it is from nuclei at high scattering angles. (b) A part of an experimental UED pattern showing multiple Bragg reflections and superlattice reflections (SLRs). Nine sets of SLRs were selected for the measurements with their scattering angles marked by the dashed lines. (c) Intensity variation (normalized by the averaged intensity before pumping) vs. time delay measured from representative SLRs. The solid lines are fittings of the experimental data to a two-exponential model.

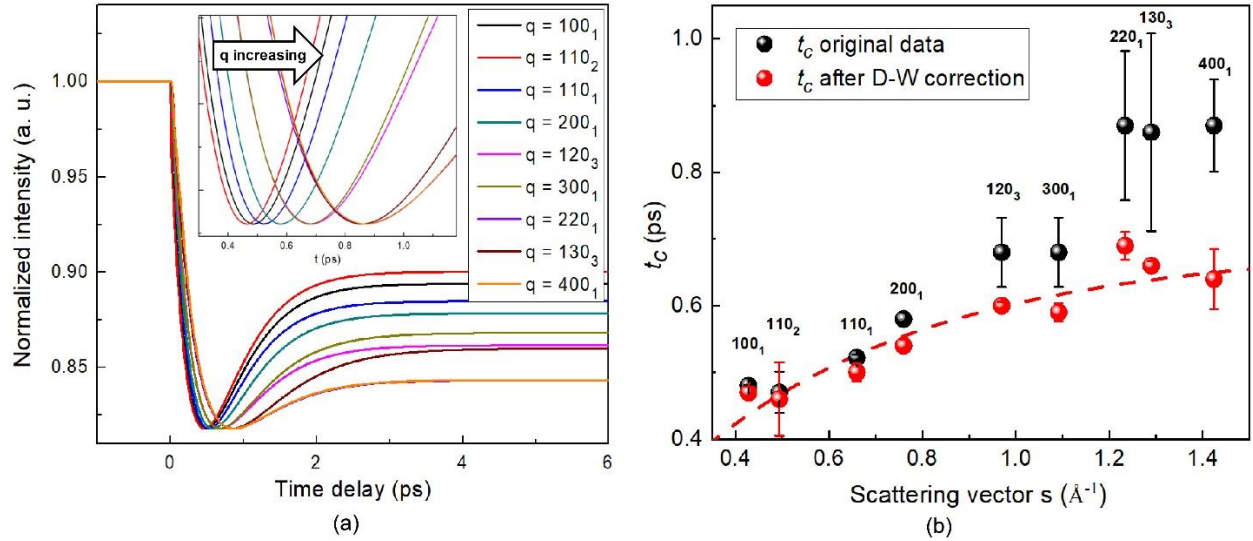

Figure 2. Measurement of the time at the cusp point $t_c$ of the SLRs during the lattice relaxation process as a function of the scattering vector length $s$. (a) Two-exponential fitting curves for nine sets of SLRs, showing a shift of $t_c$ toward increased time delay with the increase of scattering vector **s**. (b) A plot of $t_c$ vs. the scattering vector length $s$ of the SLRs. The black dots are measured directly from the fitting of the raw diffraction data. The red dots are the measurements after the removal of the influence of the Debye-Waller (D-W) factors on the values of $t_c$ (see supplementary materials for the details of the D-W correction). The red dashed line is a guide to the eye for the red dots. The error bars mean the deviation of the $t_c$ values determined in two distinct fitting methods (see supplementary Figures S2, S2-1 and S2-2).

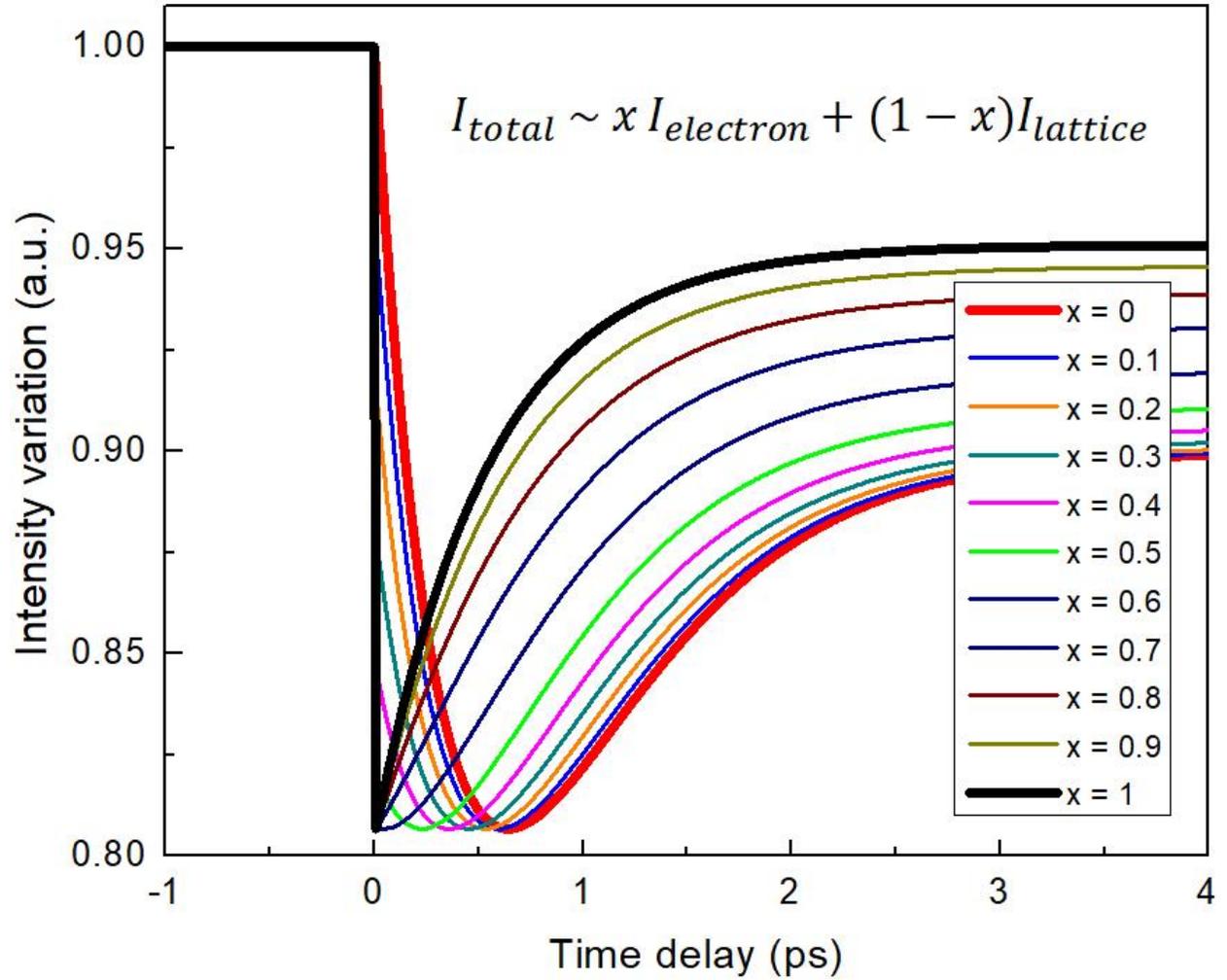

Figure 3. Intensity variation as a function of time derived from dynamic models. The bold black curve reflects the electron while the bold red curve reflects the lattice dynamics considering the electron relaxation rate and the electron-lattice coupling time constant. Colored curves are the linear combinations of the black and the red curves, indicating a shift of $t_c$ as a function of x, or the weight of the electron dynamics, in the total measurement.

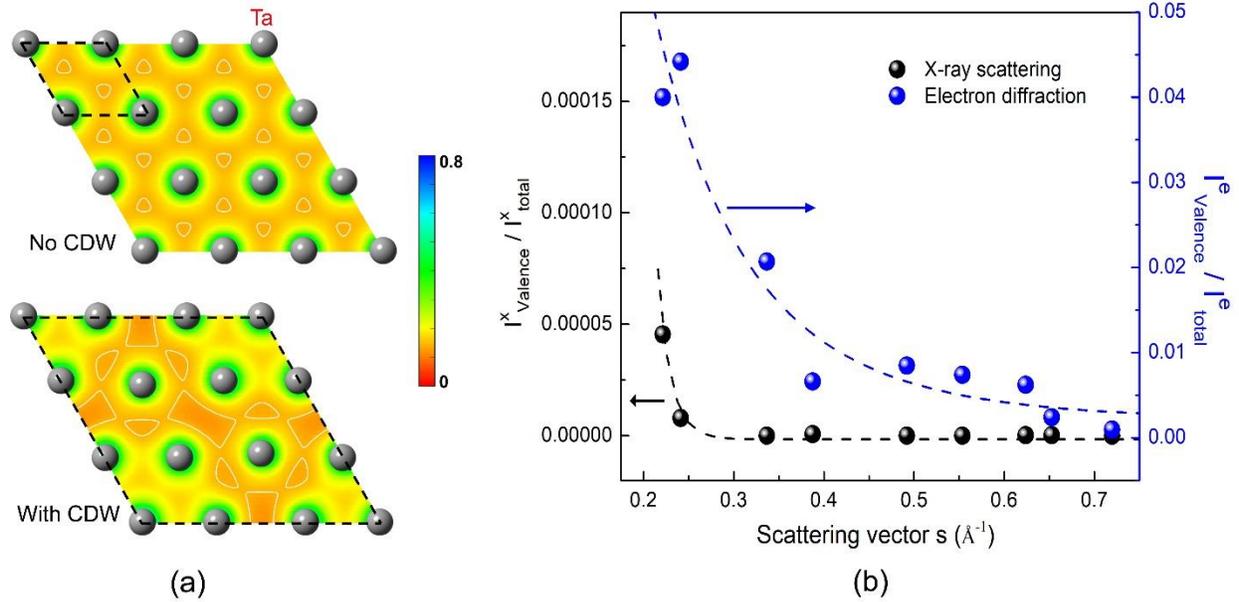

Figure 4. Calculated (planar) charge density distributions and associated x-ray/electron scattering intensities. (a) Charge density mapping of the Ta plane (Ta atoms are shown) in 1T-TaSe$_2$ based on DFT calculations for the state with no CDW (upper; unit cell marked by the black dashed lines) and the CDW state (lower; unit cell is 3×3 times larger). The color scale shows the number of electrons in unit-cell volume. (b) Structure factors of the CDW state were calculated, corresponding to the nine measured SLRs, using valence charges (5$d$ and 6$s$ electrons for Ta atoms and 4$p$ electrons for Se atoms) and total charges of Ta and Se atoms based on the electronic structures in (a). The ratios of the scattering intensity from valence electrons to the total intensity are obtained and plotted for x-ray (black dots) and electron diffraction (blue dots). Note that the scales are different for x-ray and electron scattering. The dashed lines are guides to the eye.

# Supplementary Materials

# Concurrent probing of electron-lattice dephasing induced by photoexcitation in 1T-TaSeTe using ultrafast electron diffraction

Jun Li[1], Junjie Li[1], Kai Sun[2], Lijun Wu[1], Renkai Li[3], Jie Yang[3], Xiaozhe Shen[3], Xijie Wang[3], Huixia Luo[4], Robert J. Cava[4], Ian K. Robinson[1,5], Xilian Jin[1,6], Weiguo Yin[1], Yimei Zhu[1], Jing Tao[1, *],

[1]Condensed Matter Physics & Materials Science Department, Brookhaven National Laboratory,

Upton, NY 11973, USA

[2]Department of Physics, University of Michigan, Ann Arbor, MI 48109, USA

[3]SLAC National Accelerator Laboratory, Menlo Park, CA, USA.

[4]Department of Chemistry, Princeton University, Princeton, NJ 08544, USA

[5]London Centre for Nanotechnology, University College, London WC1E 6BT, UK

[6]College of Physics, Jilin University, Changchun, Jilin 130012, P.R. China

**Material information, experimental settings and UED data analysis.**

The 1T-TaSeTe single crystals were grown by chemical vapor transport with iodine as a transport agent [26]. The single crystalline foils with an average thickness around 50 nm and diameter larger than 300 μm were mechanically exfoliated from the bulk using adhesive tape, and transferred to the nickel TEM meshes with the aid of acetone.

The experiments reported here were performed on the 3.5-MeV-UED setup at the Stanford Linear Accelerator Laboratory which can achieve 100 fs temporal resolution. Using this experimental setup, the free-standing foils were uniformly excited by a 60-fs [full width at half maximum (FWHM)] laser pulse with a photon energy of 1.55 eV (center wavelength of 800 nm) and a fluence of 3.5 mJ/cm$^2$. The optical pulses at a repetition rate of 120 Hz were focused down to 1.5 mm on the sample with uniform thickness to trigger electron and lattice dynamics. The laser pump pulse and probing electron beam are colinear in the UED instrument. The excited states within an illuminated area of diameter 400 μm were probed by well-synchronized 3.5 MeV electron pulses containing ~ 10$^5$ electrons, ensuring a large number of accessible reflections were recorded in each diffraction pattern. The sample temperature of 26 K was achieved using a conducting sample holder cooled by liquid helium. The selected area electron diffraction patterns included in the Supplementary Figure S3 (b) and (c) were obtained with a JEOL ARM 200F microscope operating at 200 kV @ BNL.

At each delay time, 10 patterns were acquired under the same condition and integrated to improve the signal-to-noise ratio. 94 patterns with different delay times were obtained in each full scan. This process was repeated ten times and then the ten sets of data were added up after shift correction at each delay time to minimize the systematic errors and improve data quality. For intensity measurements, the superlattice reflection peaks were fitted by a two-dimensional rotated Gaussian function

$$FIT_\mathbf{H} = A\exp\left\{-\frac{[(x-x_0)\cos\theta-(y-y_0)\sin\theta]^2}{2\sigma_1^2} - \frac{[(x-x_0)\sin\theta+(y-y_0)\cos\theta]^2}{2\sigma_2^2}\right\} + (Bx + Cy + D)$$

, where $(Bx+Cy+D)$ is used to estimate the local background intensity. The obtained value $I_\mathbf{H} = FIT_\mathbf{H} - (Bx + Cy + D)$ is regarded as the measured intensity of diffraction spot.

**A comparison between kinematic and dynamic simulations of the UED patterns.**

We performed both dynamic and kinematic electron diffraction simulations. After compared with experimental data, we use kinematic simulations to extract atomic positions because the dynamic simulations show only minor perturbations to the kinematic simulations in this case. We demonstrate that the kinematic treatments are suitable for this case mainly because of two factors. The first is the extremely high energy of incident electrons, i.e., 3.5 MeV, which leads to less multiple scattering [35]. The second is the well-known rippling of the thin sample foil within the large electron illuminating area (~300 $\mu$m as the diameter), leading to a similar effect as precession electron diffraction [36]. The detailed methods are shown in Figure S3 and in ref. [27].

Based on the time-dependent atomic positions extracted in ref. 27, diffraction simulation results show that $t_c$ of all the SLRs should be the same under the effect of lattice distortions. Simulated intensity variations for reflection $(100_1)$ and $(400_1)$ are measured and plotted in Fig. S4 (kinematic simulations of electron diffraction patterns as a function of time delay) and Fig. S5 (dynamic simulations of electron diffraction patterns as a function of time delay).

For the Bloch wave method, the time-dependent intensities of $(100_1)$ and $(400_1)$ spots are calculated with the sample thickness of 10 nm and the accelerating voltage of 3.5 MeV, shown in Figure S5. To emphasize the impact of structural parameters on the cusp positions, all the Debye-Waller factors are set to 0 in simulations. In order to objectively select the electron beams included in calculations, the five beam selection criteria are chosen as $|\boldsymbol{g}|_{\max} = 3.5$ Å$^{-1}$, $|2KS_g|_{\max} = 4.0$ Å$^{-2}$, $|U_g/2KS_g|_{\min} = 0.01$, $|2KS_g|_{\min} = 3.0$ Å$^{-2}$ and $\omega_{\max} = 35$. The strong and weak beams are classified with the structural parameters at $\Delta t = -2$ ps, and then the classification does not change over time. The simulated result indicates that the cusp positions at different scattering angles remain the same even though the dynamic scattering effect is taken into consideration.

As a result, we found that multiple electron-scattering effects, which redistribute the intensities among the reflections, do not affect the determination of the cusp point in our measurements. During photoexcitation, it is reasonable to consider that the sample thickness and orientation remain the same, especially in short time delay manifested in this manuscript. Consequently, atomic-position changes are more decisive for the reflection intensity variations than multiple scattering effects wherein the dynamic situations vary as higher orders of small atomic displacements.

On the other hand, the dynamic scattering effects is very important for general consideration using the method demonstrated here. Kinematic scattering treatments of the UED results may not be suitable for conventional low voltage UED experiments and/or some other UED samples, which depend on the materials' dimensionality and sample preparations.

**A comparison between DFT calculations using different types of CDW structures.**

Due to the difficulty in constructing incommensurate structures and distinguishing Se and Te elements in this material, the DFT calculations (results shown in Fig. 4) were performed with restrictions to find a structure in 1T-TaSe$_2$ with local energy minimum that is associated with the (3 × 3; commensurate) CDW, which is experimentally observed in 1T-TaSeTe sample at 26 K but with incommensurate CDW reflections (q is not exactly 1/3). Note that the low temperature phase of 1T-TaSe$_2$ shows diffraction patterns with the a Star of David CDW ( (SoD; with a $\sqrt{13} \times \sqrt{13}$ unit cell and reflections at commensurate positions). Nevertheless, 1T-TaSe$_2$ material was reported to have (3 × 3) incommensurate CDW at high temperatures [9], that could justify our DFT calculations with local minimum that result in the (3 × 3) CDW structure.

We further performed additional DFT calculations that result in the charge density map with SoD CDW ($\sqrt{13} \times \sqrt{13}$). Results are shown in Figure S7. The arrangement of the results and plots are similar to the content in Fig. 4. We use the same amount of valence electrons as the calculations done in the (3 × 3) CDW to calculate the weight of the intensities from valence electrons in the total intensities, as a function of scattering angle. It is evident that, despite the type of the CDW [ the (3 × 3) CDW or the SoD CDW], electron scattering is very powerful to distinguish valence electrons from the total charge and nucleus - the intensity weight from valence electrons not only is much higher than that from x-ray but also can be measured at higher scattering angles.

**Supplementary figures**

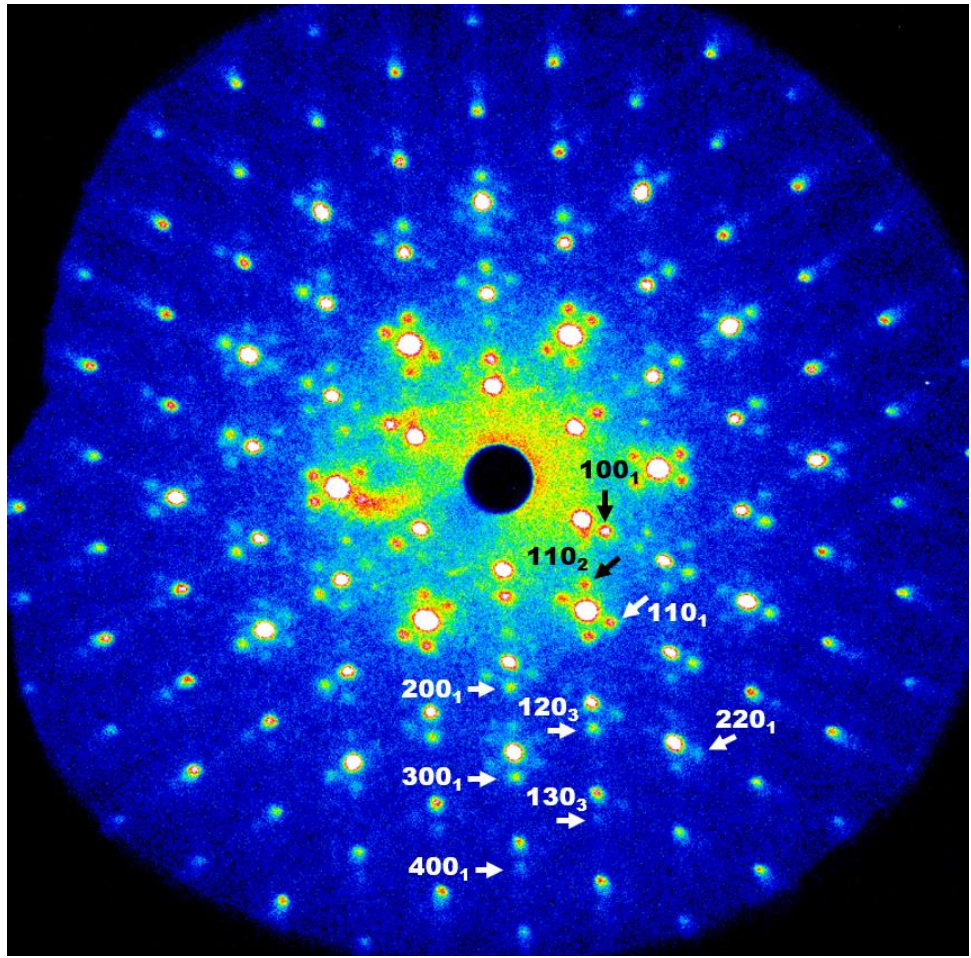

Figure S1. A typical UED pattern obtained before time zero with all the Bragg peaks and superlattice reflections (SLRs) presented. The nine selected SLRs are indexed with arrows.

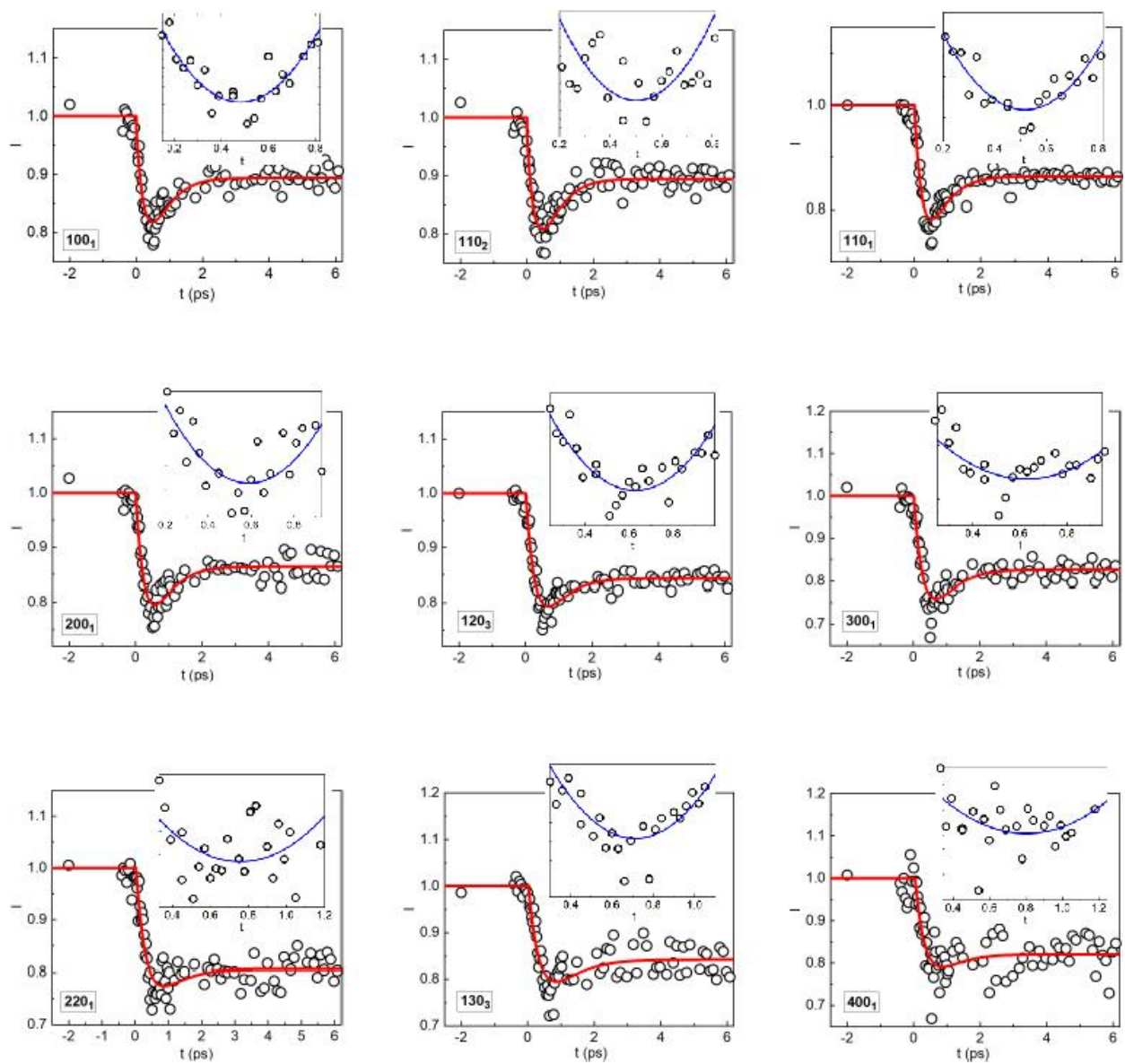

Figure S2. Measured normalized intensities ($\frac{I(t')}{I(t_0)}$, where $I(t_0)$ is the average measurement before time zero) as the vertical plots of nine selected SLRs. The horizontal plots are the time delay in ps. The red lines are the curve fittings using a two-exponential function. The insets show the fitting results using a simple "harmonic" function $a + b(t-t_c)^2$ (blue lines) for the data points within a narrow time window around the experimental minimum.

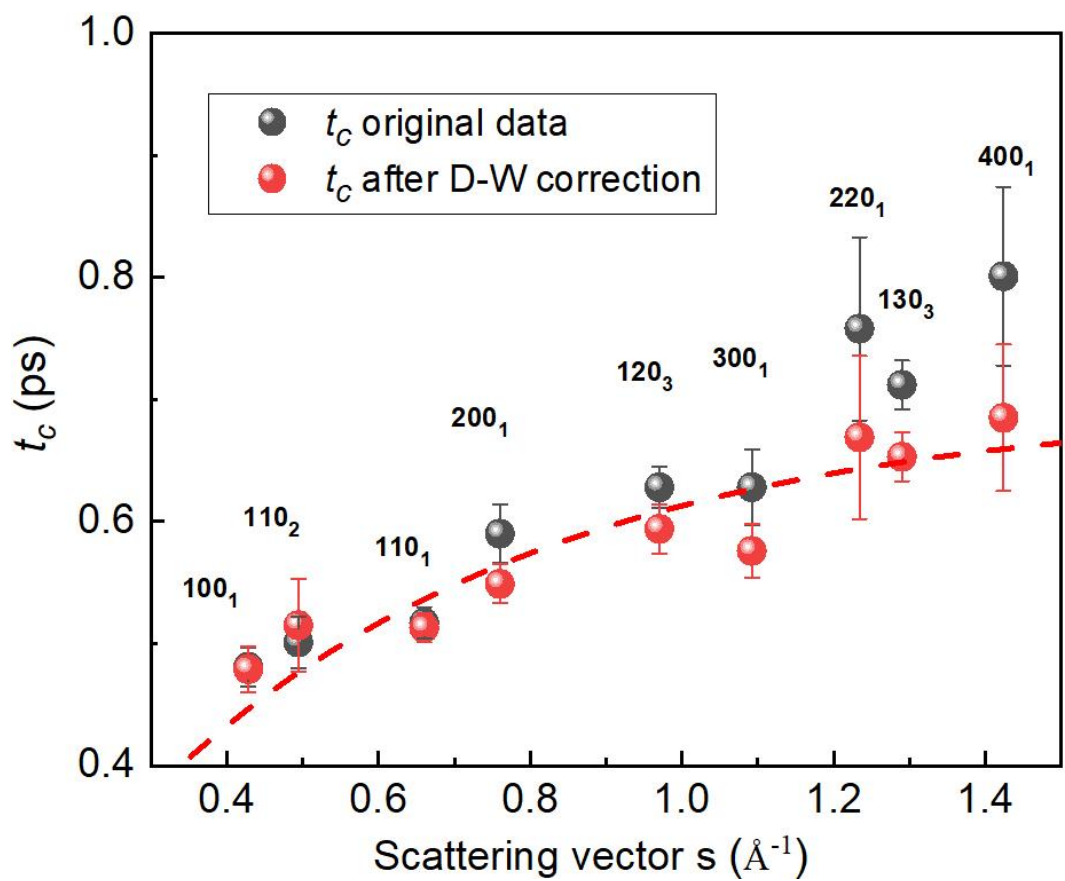

Figure S2-1. The scattering vector dependence of the cusp point $t_c$ obtained from fitting with the harmonic form (see Figure S2). The $t_c$ measurements are in quantitative agreement with the results obtained from fitting with the two-exponential fitting method (see Figure 2). Dash red line is an eye guide.

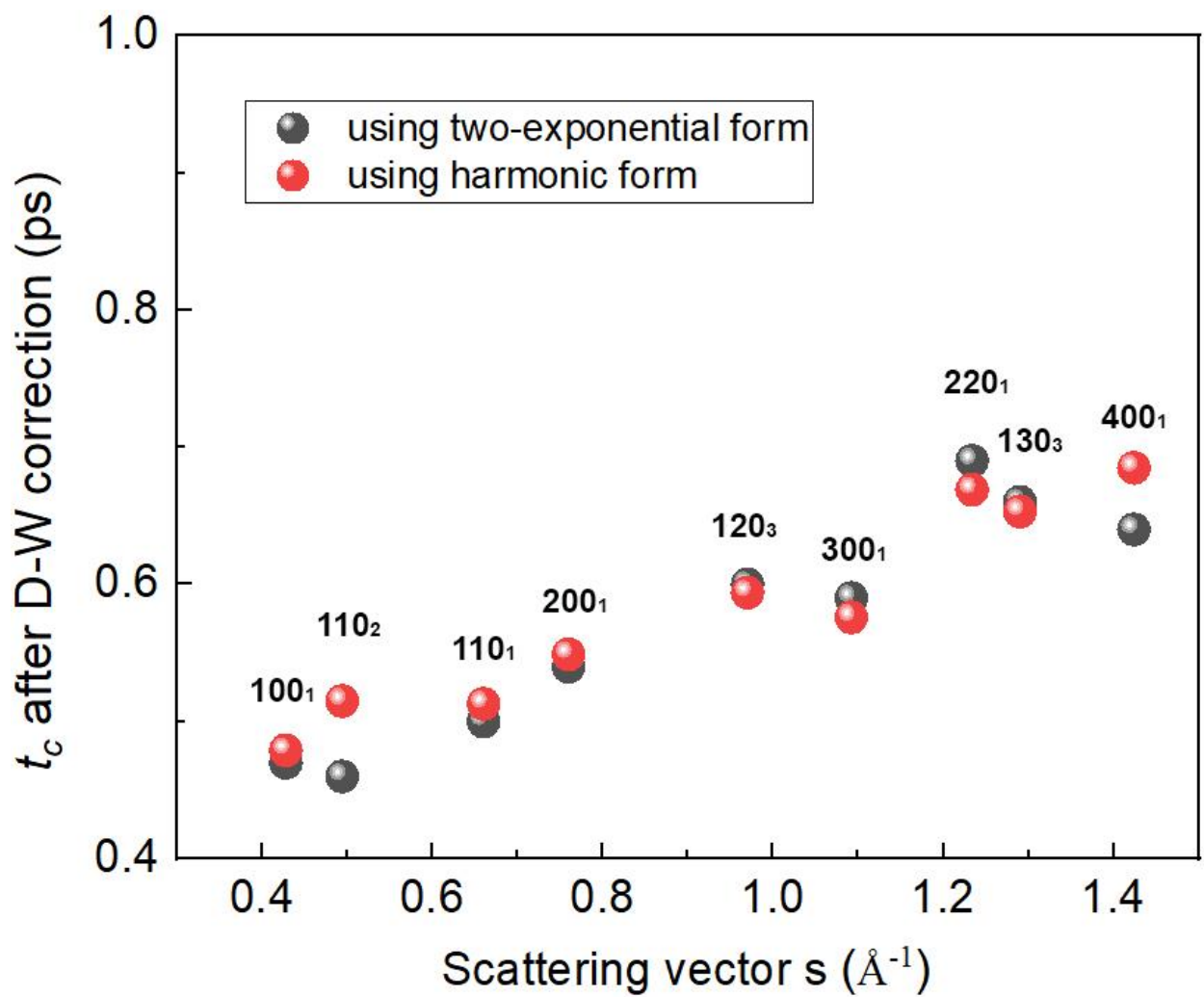

Figure S2-2. A comparison of $t_c$ extracted from data after D-W correction using two-exponential fitting form and harmonic fitting form. The error bars of red dots plotted in Fig. 2b in the main text are calculated by the difference between black and red sets here.

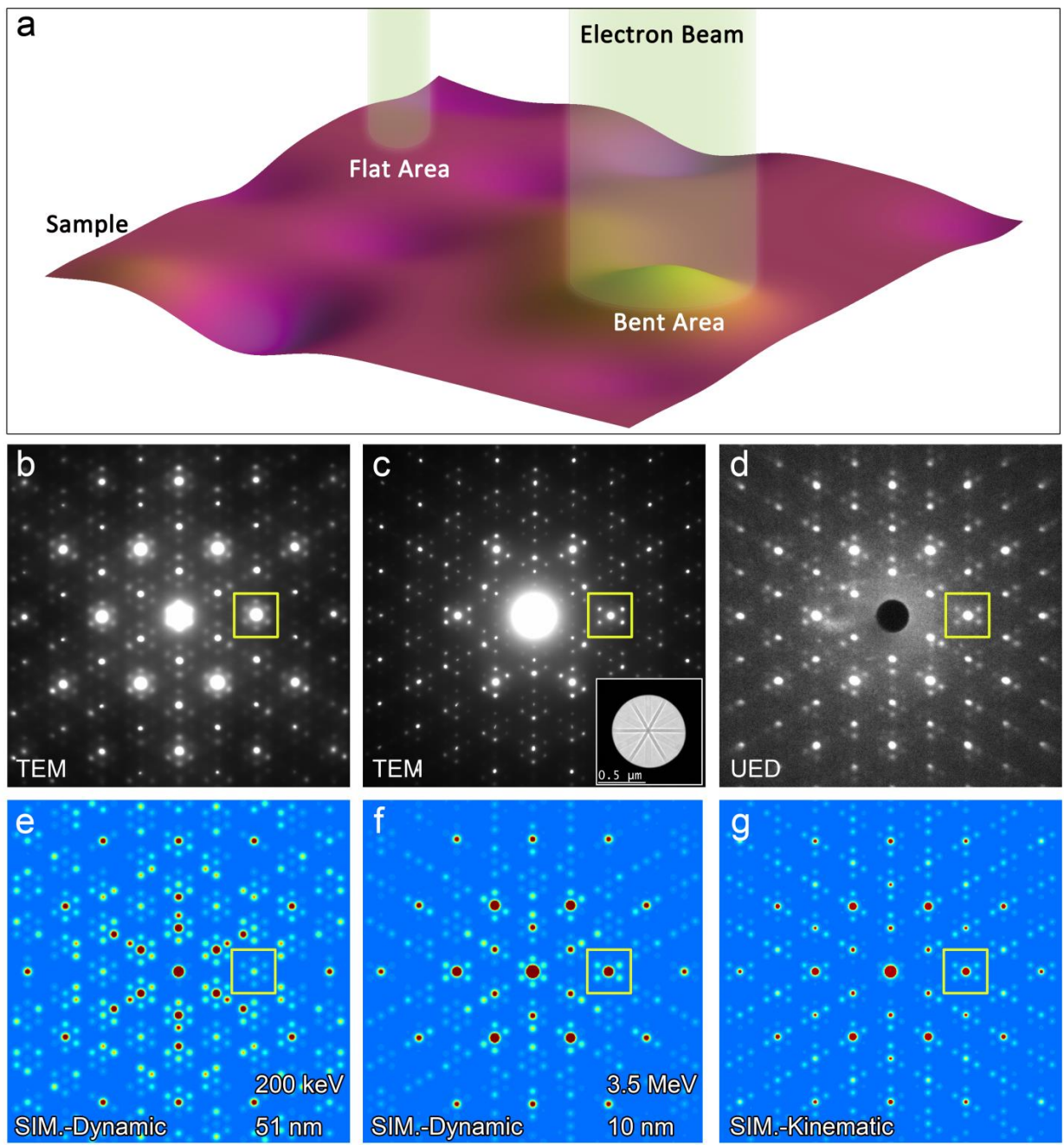

Figure S3. (a) A schematic drawing to show the UED sample morphology and electron diffraction acquisition. (b) TEM diffraction pattern acquired from a flat area of the TaSeTe sample demonstrated in (a), with the area diameter ~ 100 nm. Six CDW satellites around the (110) spot (within the yellow box) are at about the same intensity due to multiple electron scattering (dynamic diffraction effects). (c) A TEM diffraction pattern acquired from a rippled region of the same sample demonstrated in (a), with the area diameter ~ 700 nm. The inset shows the corresponding bright-field electron diffraction contrast image of the region with a symmetric bending contour (i.e., the diffraction obtained from the entire rounded valley or protuberance, the circle indicates

the position of selective area aperture). Four of the six satellites are much brighter than the other two. Because the sample area bends symmetrically so that the obtained electron diffraction has a similar effect as the precession diffraction technique, i.e., the ensemble of intensities in the pattern behaves in a "kinematick-like" fashion. (d) A typical UED pattern which shows similar intensity distribution as presented in c), especially the CDW satellites around the (110) spot. (e) Simulated electron diffraction pattern based on the Bloch wave theory with the thickness of 51 nm (accelerating voltage = 200 kV) that shows a good fitting to the experimental pattern (obtained at 200 kV) in (a). The intensities of the six satellite spots are almost equal, indicating that the dynamic diffraction effect can result in this intensity feature. (f) Simulated electron diffraction pattern based on the Bloch wave theory with the thickness of 10 nm (accelerating voltage = 3.5 MeV). (g) Simulated electron diffraction pattern based on the kinematic theory. A comparison between the dynamic simulation in (f) and the kinematic simulation in (g), using characteristics in the yellow box, shows similar intensity distributions in the patterns. Indeed, the dynamic scattering in our case only adds minor perturbation in intensity derived from the kinematic simulation, which qualitatively matches well with the experimental patterns shown in (c) and (d).

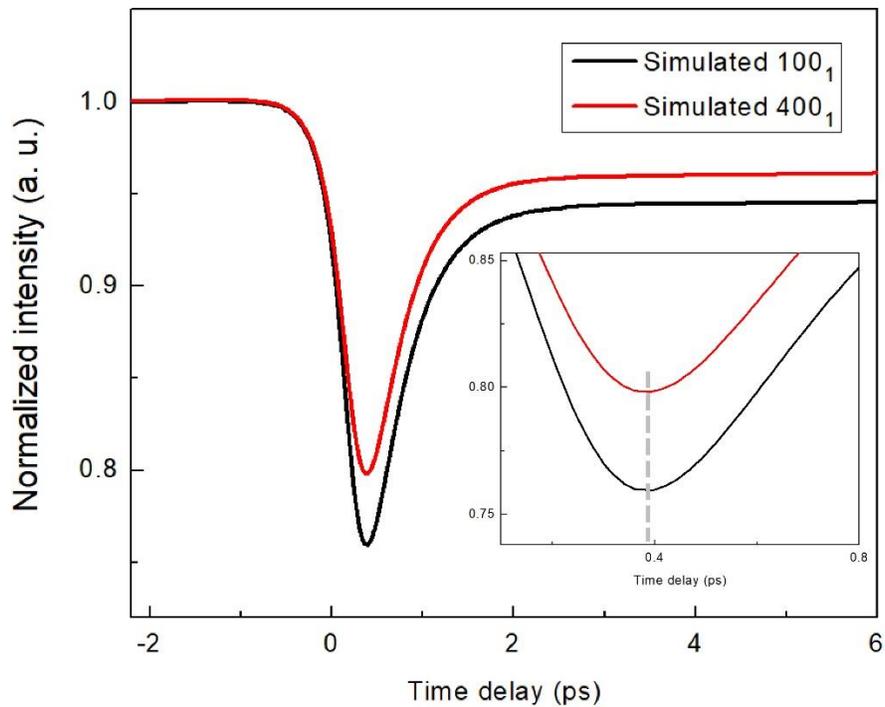

Figure S4. Time-dependent intensities for reflection $(100_1)$ and $(400_1)$ using kinematic electron diffraction simulations based on the time-dependent atomic-positions extracted in ref. 27. It is evident that, without other considerations, the same set of atomic positions results in the same time at the cusp (enlarged in the inset) for reflections at different scattering angles.

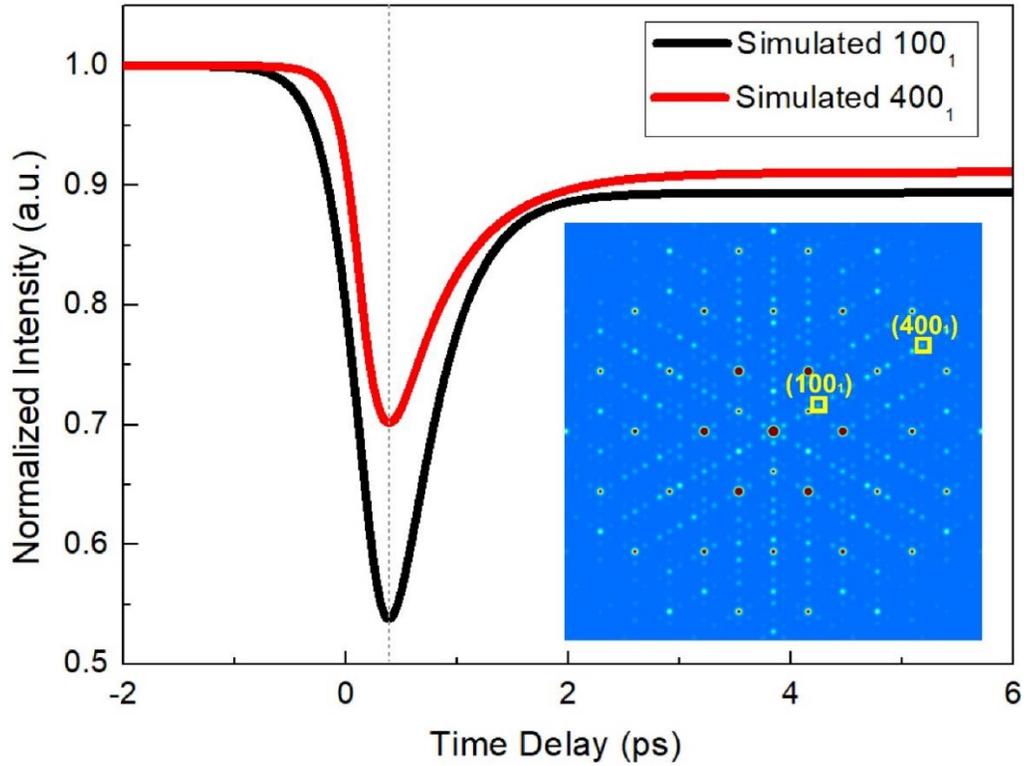

Figure S5. Using the Bloch wave method at the sample thickness of 10 nm, intensity variations of $(100_1)$ and $(400_1)$, two CDW superlattice reflections, as a function of time delay measured from dynamically simulated UED patterns based on the time-dependent atomic-positions extracted in ref. 27. One simulated UED pattern is shown in the inset. The cusp points for both reflections take place at a time delay ~ 400 fs, which is consistent with the measurements from kinematic simulations of the UED patterns.

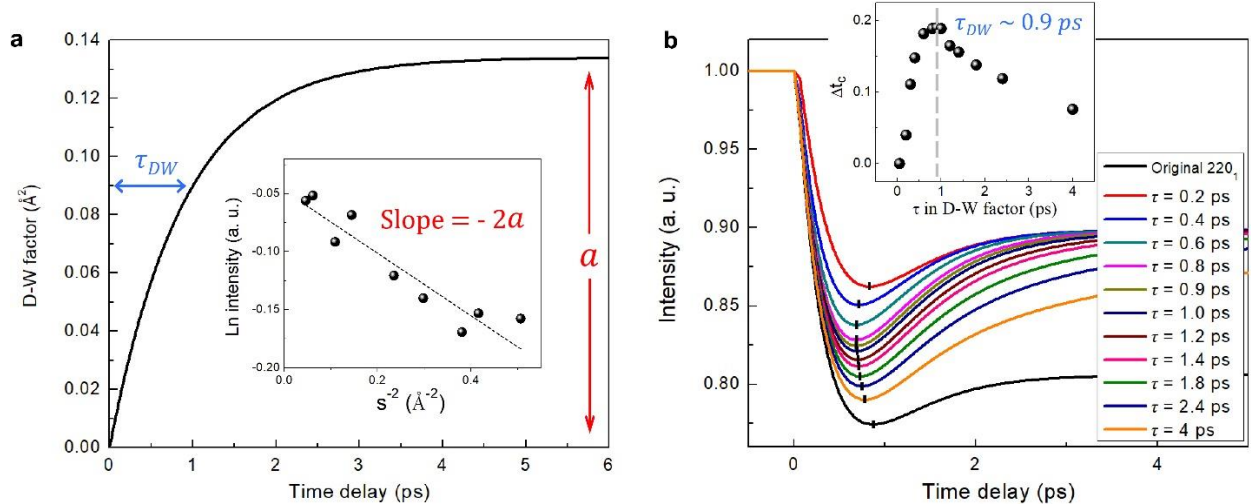

Figure S6. The determination of the D-W factors. Assuming that the D-W factor is expressed to be $B(t') = a(1 - e^{-t'/\tau_{DW}})$, which is plotted in (a), the value of parameter $a$ can be obtained by measuring the slope of $\ln\left[\frac{I(t')}{I(t_0)} / \frac{I(u')}{I(u_0)}\right]$ vs. $s^2$ (i.e., $-2B(t')s^2$ vs. $s^2$) at $t'$ at 6 ps (data is shown in Supplementary Table S1), which is determined to be ~ 0.134 (Å$^2$). The effect of D-W factor on the cusp time is illustrated in b) using different time constant $\tau_{DW}$. Taking the experimental intensity measured from $220_1$ reflection (the black curve in (b) at the bottom) as an example, it is seen that the plots with the D-W effect removed from the experimental measurements (i.e., $\frac{I(t')}{I(t_0)} / e^{-2B(t')s^2}$) shows a different change at the cusp time ($\Delta t_c$) as a function of the time constant $\tau_{DW}$. The maximum D-W effect (most $\Delta t_c$) takes place at $\tau_{DW}$ ~ 0.9 ps, which gives the upbound limit of the D-W effect in our consideration.

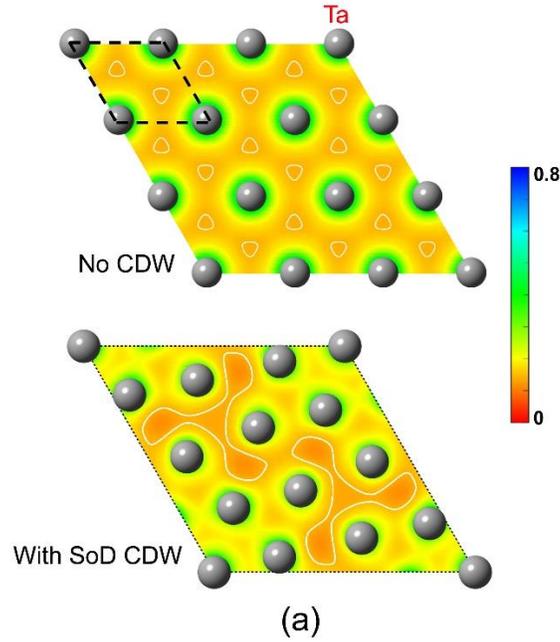

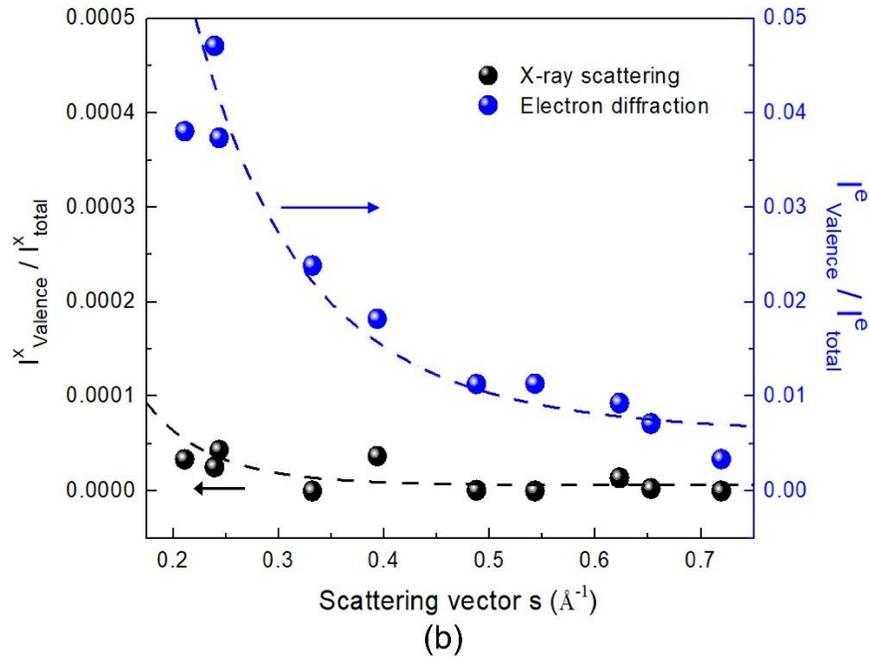

Figure S7. Calculated (planar) charge density distributions and associated x-ray / electron scattering intensities. Note that this figure is similar to Fig. 4 in the main text but using charge density in a Star of David (SoD) CDW structure. (a) Charge density mapping of the Ta plane (Ta atoms are shown) in 1T-TaSe$_2$ based on DFT calculations for the state with no CDW (upper; unit cell marked by the black dashed lines) and the CDW state (lower; unit cell is $\sqrt{13} \times \sqrt{13}$ times larger; a whole unit cell is not shown for a comparison to Fig. 4 in the main text). The color scale shows the number of electrons in unit-cell volume. (b) Structure factors of the CDW state were calculated, corresponding to the nine measured SLRs, using valence charges ($5d$ and $6s$ electrons for Ta atoms and $4p$ electrons for Se atoms) and total charges of Ta and Se atoms based on the electronic structures in (a). The ratios of the scattering intensity from valence electrons to the total intensity are obtained and plotted for x-ray (black dots) and electron diffraction (blue dots). Note that the scales are different for x-ray and electron scattering. The dashed lines are guides to the eye.

Supplementary Table S1. Experimental measurements ($\frac{I(t')}{I(t_0)}$) and electron diffraction simulations ($\frac{I(u')}{I(u_0)}$) for nine selected SLRs.

| Index of SLR | $s^2$ (Å$^{-2}$) | Measured $\frac{I(t')}{I(t_0)}$, $t'$ at 6 ps | Simulated $\frac{I(u')}{I(u_0)}$, $t'$ at 6 ps | $\frac{I(t')}{I(t_0)} / \frac{I(u')}{I(u_0)}$ $= e^{-2B(t')s^2}$ (), $t'$ at 6 ps | $\ln\left[\frac{I(t')}{I(t_0)} / \frac{I(u')}{I(u_0)}\right]$ $(= -2B(t')s^2)$, $t'$ at 6 ps |
|---|---|---|---|---|---|
| 100$_1$ | 0.04574 | 0.89398 | 0.94552 | 0.94549 | -0.05605 |
| 110$_2$ | 0.06118 | 0.89476 | 0.94209 | 0.94976 | -0.05155 |
| 110$_1$ | 0.10883 | 0.86204 | 0.9447 | 0.9125 | -0.09156 |
| 200$_1$ | 0.14435 | 0.86378 | 0.9249 | 0.93392 | -0.06837 |
| 120$_3$ | 0.23501 | 0.84319 | 0.95117 | 0.88647 | -0.12051 |
| 300$_1$ | 0.2981 | 0.82627 | 0.95047 | 0.86933 | -0.14004 |
| 220$_1$ | 0.38082 | 0.80583 | 0.95444 | 0.8443 | -0.16925 |
| 130$_3$ | 0.41633 | 0.8421 | 0.98136 | 0.85809 | -0.15304 |
| 400$_1$ | 0.50699 | 0.82049 | 0.96058 | 0.85416 | -0.15764 |

**Conversion between Electron and X-ray structure factors.** The electron ($f^{(e)}(s)$) and X-ray ($f^{(X)}(s)$) atomic scattering factors at scattering vector magnitude s are related by the Mott formula:

$$f^{(e)}(s) = \frac{|e|}{16\pi^2\varepsilon_0} \frac{[Z-f^{(X)}(s)]}{s^2} \quad (1)$$

where $e$ is the electron charge, $\varepsilon_0$ is the vacuum permeability and $Z$ is the atomic number of the atom. By definition, the electron structure factor $F_g^{(e)}$ can be written as

$$F_g^{(e)} = \sum_j f_j^{(e)} \exp(-2\pi i \boldsymbol{g} \cdot \mathbf{r}_j) \quad (2)$$

where the summation extends over all atoms in the illuminating area. Using Eq. (1), Eq. (2) can be written in the form

$$\begin{aligned}
F_g^{(e)} &= \sum_j \frac{|e|}{16\pi^2\varepsilon_0 s^2}\left[Z_j - f_j^{(x)}(s)\right]\exp(-2\pi i \boldsymbol{g} \cdot \mathbf{r}_j) \\
&= \frac{|e|}{16\pi^2\varepsilon_0 s^2}\left[\sum_j Z_j \exp(-2\pi i \boldsymbol{g} \cdot \mathbf{r}_j) - \sum_j f_j^{(x)}(s)\exp(-2\pi i \boldsymbol{g} \cdot \mathbf{r}_j)\right] \\
&= \frac{|e|}{16\pi^2\varepsilon_0 s^2}\left[\sum_j Z_j \exp(-2\pi i \boldsymbol{g} \cdot \mathbf{r}_j) - F_g^{(x)}\right]
\end{aligned}$$
(3)

where the X-ray structure factor is defined in its usual form $F_g^{(x)} \equiv \sum_j f_j^{(x)} \exp(-2\pi i \boldsymbol{g} \cdot \mathbf{r}_j)$. The X-ray structure factors calculated from DFT can thus be converted to the electron structure factors by Eqn. (3).